\documentclass[preprint,superscriptaddress]{revtex4-2}
\usepackage{amsmath,amssymb}
\usepackage{hyperref}
\usepackage{blindtext}
\usepackage{graphicx}
\usepackage{xcolor}
\graphicspath{ {.} }
\usepackage{comment} 
\usepackage{subfigure}
\usepackage{color}
\usepackage{tikz}
\usepackage{graphicx}
\usepackage{float}
\usepackage{todonotes}
\usepackage[a4paper]{geometry}
\usepackage{float} 

\begin{document}
	\makeatletter
\renewcommand{\@fnsymbol}[1]{\ifcase#1\or *\or \#\or $\dagger$ \else\@arabic{#1}\fi}
\makeatother

\title{ Emergent Rotational Order and Re-entrant Global Order of Vicsek Agents in a Complex Noise Environment }
\affiliation{School of Materials Science and Engineering, Harbin Institute of Technology (Shenzhen), Shenzhen, Guangdong, 518055, China}
\affiliation{Department of Physics, Indian Institute of Technology (Delhi), Delhi, India, 202002, India}

\author{Mohd Yasir Khan$^{1,2}$}
\email{physics.yasir@gmail.com}

\thanks{\\This work was conducted as part of the Ph.D. work at IIT Delhi by the author.}

\begin{abstract}\label{Labstract}
Noisy pursuit in complex environments drives emergent collective behaviors in active matter systems. A compelling platform to study the impact of environment cues is provided by the standard Vicsek model for studying flocking and swarming phenomena. In this study, we explore the collective dynamics of Vicsek agents in a complex noise environment, featuring a noiseless circular region (\(\eta_{\text{c}} = 0.0\)) surrounded by a noisy outer region (\(\eta_{\text{b}} = 1.0\), tunable), with a mutually repelling interactions. By varying the outer noise intensity, we observe an emergent rotational order (\(\phi_r\)) that peaks at higher noise levels (\(\eta_{\text{b}}\sim 1\)), as revealed by phase and susceptibility plots. 
Global order follows ($\phi$) follows a `U' shaped curve, $\phi \sim 0.965$ at $\eta_b=0$, dies down to $\phi \sim 0.57$ at $\eta_b=0.9$ and re-enters at $\eta_b > 1$ and peaks $\phi \sim 0.960$ at $\eta_b=1.5$. The latter rise attributing to $\phi_r$ increase.
Higher particle velocities enhance escape rates (\(\kappa\)) from the circular region, with slower-moving agents exhibiting greater virtual confinement. We quantify escape dynamics through time-averaged and first-passage escape rates, demonstrating velocity-dependent retention on the probability of finding the bi-motility agent flocks at a give time resulting in segregation and trapping. Introducing a gradual noise increase from the circle's center to the outer region reduces both global (\(\phi\)) and rotational (\(\phi_r\)) order, underscoring the impact of environmental heterogeneity and sudden annealing over gradual change. These findings offer insights into predicting and manipulating active agent dynamics in heterogeneous environments, with applications in biological and synthetic swarming systems.
\end{abstract}
\maketitle

\section{Introduction}\label{LIntro}
Noisy pursuit in active matter systems, where self-propelled agents navigate complex environments, has emerged as a vibrant area of research \cite{chatterjee2025noisy,frey2005brownian,ramaswamy2010mechanics}. Noise introduces randomness that disrupts group cohesion, yet paradoxically shapes mesmerizing collective behaviors such as flocking, swarming, schooling, herding, segregation, and clustering \cite{jhawar2020noise,yatesa2009inherent,yates2009united}. These phenomena are ubiquitous across scales, from bird flocks and fish schools to bacterial colonies and synthetic robotic swarms, driven by local interactions and environmental cues both in and out of the presence of heterogeneity \cite{couzin2005effective,king2012murmurations,mathijssen2019collective,peleg2018collective,zhang2021collective}. Natural environments, characterized by crowded boundaries \cite{khan2022effect}, chemical gradients \cite{peng2025self}, topological constraints \cite{mackay2025emergent}, or curved surfaces \cite{xiao2018review}, introduce spatial complexity that profoundly influences these dynamics.

In passive systems, noise is essential for crossing thermodynamic barriers, to facilite advance material modeling \cite{varma2024breaking}, and in synthetic systems, such as active colloids, noise can be tuned via temperature modulation or magnetic fields controlling rotational diffusion timescales, enabling precise manipulation of structure, dynamics, and functionality \cite{lowen2025colloidal,evers2013colloids,frechette2025active,simmchen2024active}.  In natural systems, noise arises from diverse sources: wind turbulence, sensing errors, and individual behaviors in bird flocks \cite{couzin2023benefits}; water currents, predator avoidance, and sensory noise in fish schools \cite{king2012murmurations}; thermal fluctuations, chemotaxis errors, cell variability, and hydrodynamic fluctuations in bacterial swarms \cite{mathijssen2019collective}; and air turbulence, individual decisions, fluctuating signals, and crowding in insect swarms \cite{couzin2005effective,peleg2018collective}; and  predator avoidance\cite{chakrabortty2025controlling}. These heterogeneous environments often lead to functional structures mediated by boundaries or curvature \cite{cui2025emergent,khan2022effect,mackay2025emergent,xiao2018review}.

A characteristic feature of flocking systems, is the disruption of global order under uniform noise \cite{VICSEK201271}, while spatially biased noise promotes cohesive collective behavior, enhancing global order \cite{jhawar2020noise, chan2025noise}. To statistically investigate the noisy pursuit in such complex settings, we extend the Vicsek model \cite{vicsek1995novel}, by introducing a spatially heterogeneous noise environment. Our model features a noiseless circular region (\(\eta_{\text{c}} = 0.0\), radius \(R\)) surrounded by a noisy outer region (\(\eta_{\text{b}} \leq 1.5\)), with mutually repelling interactions (\(d_{\text{min}} = 0.1\)). By varying noise intensity, velocity (\(v\)), boundary radius (\(R\)), particle number (\(N\)), interaction range (\(r_c\)), and a gradual noise gradient based on distance from the box center (\(r_{i,c}\)), we explore the emergence of global (\(\phi\)) and rotational (\(\phi_r\)) order, susceptibility (\(\chi_\phi\), \(\chi_{\phi_r}\)), and escape dynamics (\(\kappa\), \(\kappa_{\text{first}}\)).

Our key findings reveal that noise heterogeneity drives a re-entrant global order, with \(\phi\) peaking at low and high \(\eta_{\text{b}}\), and enhances rotational order, with \(\phi_r\) increasing up to 0.665 at high noise and velocity. Larger \(R\) and smaller \(N\) promote both orders, while larger \(r_c\) favors \(\phi\) over \(\phi_r\). Escape rates increase with velocity, and a bi-motility mixtures of agents separates in noisy and noiseless regions based on mobility, as seen in probability distributions at \(t = 10^4\). These results, illustrated in Figures~\ref{ffigure1}, \ref{ffigure2}, \ref{ffigure3}, and \ref{ffigure4}, reveal a re-entrant global order, enhanced rotational dynamics at high noise, and velocity-dependent escape rates, offering insights into controlling collective behavior in both natural and synthetic systems.

	\section{Simulation method} \label{SimMet}

This work incorporates the standard Vicsek model \cite{vicsek1995novel} in a complex noise environment \cite{IITDThesisyasir} with a hard-sphere boundary condition to enforce a minimum separation between agents in the 2D periodic boundary domain,  balancing alignment interactions with a spatial exclusion mechanism. $N$ self-propelled Vicsek agents are scatted at time $t=0$ randomly in a circular domain of size $R$ with $L$ being the dimension of the box with enforced periodic boundary conditions ($L=5R$). Each particle $i$ is characterized by its position $\mathbf{r}_i = (x_i, y_i)$, velocity $\mathbf{v}_i = v_0 (\cos \theta_i, \sin \theta_i)$ with constant speed $v_0$ and orientation $\theta_i$, and is subject to alignment interactions within a radius $r_c$. A boundary condition of the hard sphere ensures that the particles maintain a minimum separation distance $d_{\text{min}}=0.1$. A circular region is concentrated at the center of the box, specified with noise $\eta_{c}=0$ and the outer region with $\eta_b$. The agents are initially concentrated in the circular region.

The standard Vicsek model includes two steps in the dynamics, updating the position $ \mathbf{r}_i(t + \Delta t) = \mathbf{r}_i(t) + v_0 (\cos \theta_i, \sin \theta_i) \Delta t$ and orientation $\theta_i(t + \Delta t) = \langle\theta\rangle_i + \eta_i  $ of the particles at each time step $\Delta t$, capturing the essential flocking characteristic of agents. Here, $\eta_i = \eta  \epsilon$ is the random noise agents are subjected to, depending on the where it is located, and $\epsilon$ is drawn from a uniform random distribution $[-\pi, \pi]$. Term $\langle\theta\rangle_i$ represent the average orientation of all the agents lying in a peri-ferry to the i'th particle at a critical distance of $r_c$, with $
        \bar{\theta}_i =atan2\left( \frac{1}{n_i} \sum_{j \in \mathcal{N}_i} \sin \theta_j, \frac{1}{n_i} \sum_{j \in \mathcal{N}_i} \cos \theta_j \right),
      $
        where $n_i = |\mathcal{N}_i|$ is the number of neighbors.

We set number of particles $N = 100$, box size $R = 12$, time step $\Delta t = 0.001$, total simulation time $T = 1000$, agent speed $v_0 = 0.005$, interaction radius $r_c = 1.0$, noise amplitude $\eta_{\text{b}} = 1$ (default across the domain), $\eta_{\text{c}} = 0.0$ (within a circular region of radius $R$), and a minimum separation distance $d_{\text{min}} = 0.1$, unless stated otherwise. Periodic boundary condition is applied at the boundaries using minimum image convention.\\
At each iteration, particles are subjected to a minimum distance threshold of $d_{min}$, ensured by a hard-sphere boundary enforcement. For pairs with $r_{ij} < d_{\text{min}}$, reposition particles to achieve $r_{ij} = d_{\text{min}}$. If $\Delta \mathbf{r}_{ij} = (x_i - x_j, y_i - y_j)$ is the displacement vector, we displace the each agent using the half step $\pm \frac{d_{\text{min}} - r_{ij}}{2} \cdot \frac{\Delta \mathbf{r}_{ij}}{r_{ij}}$ iteratively until no pairs violate $r_{ij} \geq d_{\text{min}}$.

 \textbf{Order Parameter} 
        To capture the dynamical evolution of the agents we define, global translational order $ \phi = \frac{1}{N v_0} \left| \sum_{i=1}^N \mathbf{v}_i \right| $ and global rotational order $\phi_r = \frac{1}{Nv_0} \left|\sum_{i=1}^N  \hat{r}_{i,c}\times \mathbf{v}_i\right|$ parameters. Here, $r_{i,c} = (\mathbf{r}_i - \mathbf{r}_c)/ |\mathbf{r}_i - \mathbf{r}_c|$ is the relative vector of the agents from the box center $\mathbf{r}_c = (L/2, L/2)$.

\textbf{Susceptibility}
To quantify fluctuations in the order parameters \(\phi\) (global alignment) and \(\phi_r\) (rotational order), we compute the susceptibility \(\chi_{\phi} = s\frac{\langle \phi^2 \rangle - \langle \phi \rangle^2}{\langle \phi \rangle^2}\) and 
\(\chi_{\phi_r} = s\frac{\langle \phi_r^2 \rangle - \langle \phi_r \rangle^2}{\langle \phi_r \rangle^2}\) for each velocity-noise pair. 
Here, \(s=10\) is the number of ensemble for a given velocity-noise pair. 

 \textbf{Escape Event} To capture the rate of particle decay from the noiseless to noisy region, we have calculated the escape rates. We set the sampling time $\tau=1$. 
 A steady-state time-averaged escape rate $\kappa$, which measures the rate of all escape events per particle from the circle per unit time $\kappa = \frac{\sum_{0}^{T} N_{\text{escape}}(t)}{T \cdot \langle N_{\text{c}} \rangle}$, where $\langle N_{\text{c}} \rangle$ is the time-averaged number of particles inside the circle over $[0, T]$. A survival profile, $N_c(t)=N\exp(-t/\lambda)$ is fitted. To analyze the first escape events, we calculated the time-averaged first-escape rate
    $
    \kappa_{\text{first}} = \frac{\sum_{0}^{T} N_{\text{first escape}}(t)}{T\cdot \langle N_{\text{c,not escaped}} \rangle},
    $
    averaging first-escape events over $[0, T]$, normalized by the average number of unescaped particles inside the circle.

\section{Results} \label{LResults}

{\begin{figure*}[t]
    \includegraphics[width=.78\textwidth]{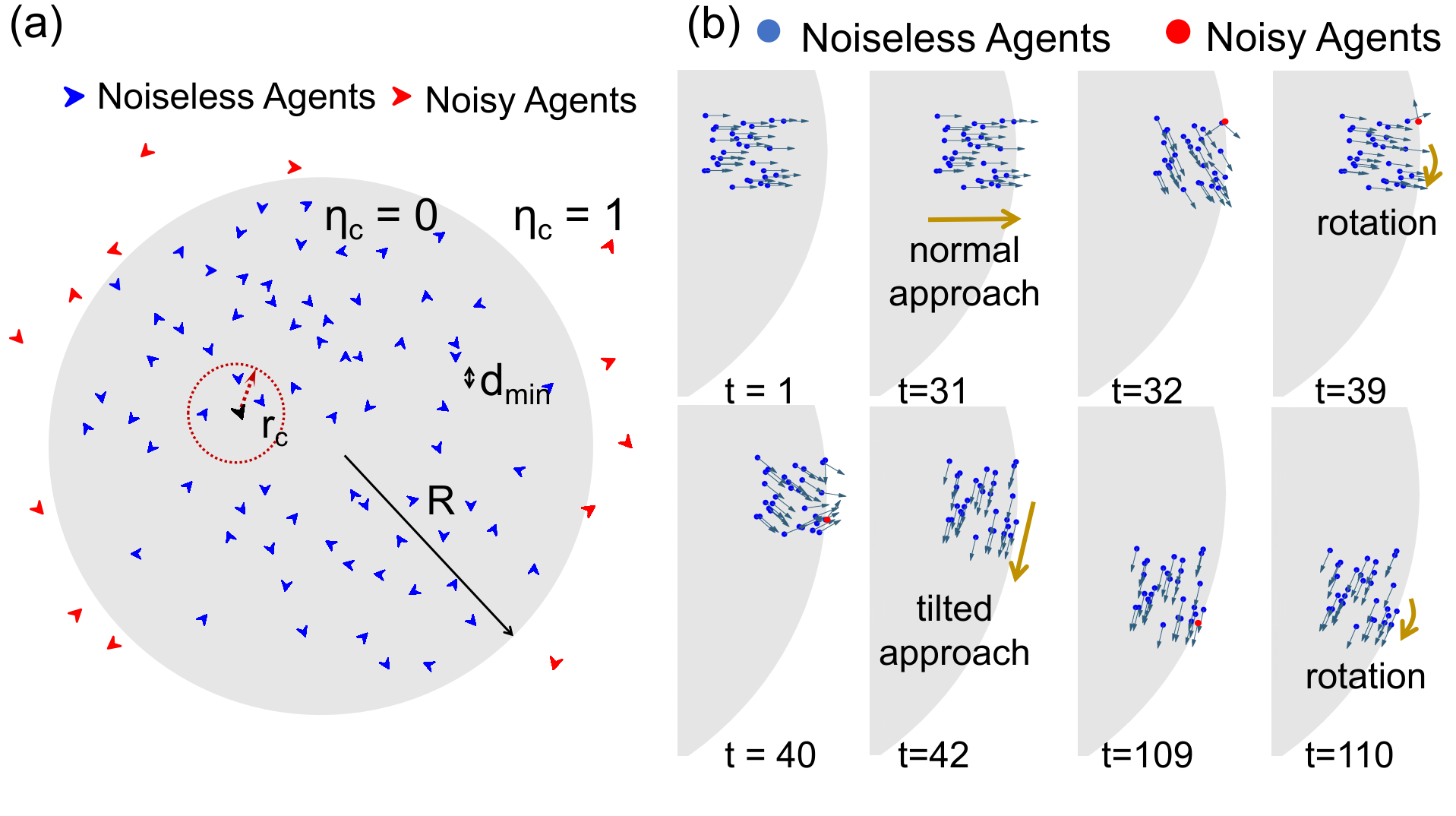}
    \caption{(a) Schematic of Vicsek agents in a complex noise environment. Agents within a circular region of radius $R$ experience zero noise ($\eta_{\text{c}} = 0.0$, blue), while those outside are subject to tunable noise ($\eta_{\text{b}} \leq 1.5$, red), interacting within a radius $r_c$. (b) Simulation snapshot showing a flock interacting with the virtual interface of the complex noise environment. At $t=31$, the flock approaches the interface normally. 
    By $t=32$, flock approaches normally to the interface and at $t=109$, flock approaches the tangentially, inducing instantaneous turbulence, driving an inward circling motion.
    }
    	\label{ffigure1}
\end{figure*}
}

\begin{figure*}[t]
    	\includegraphics[width=.88\textwidth]{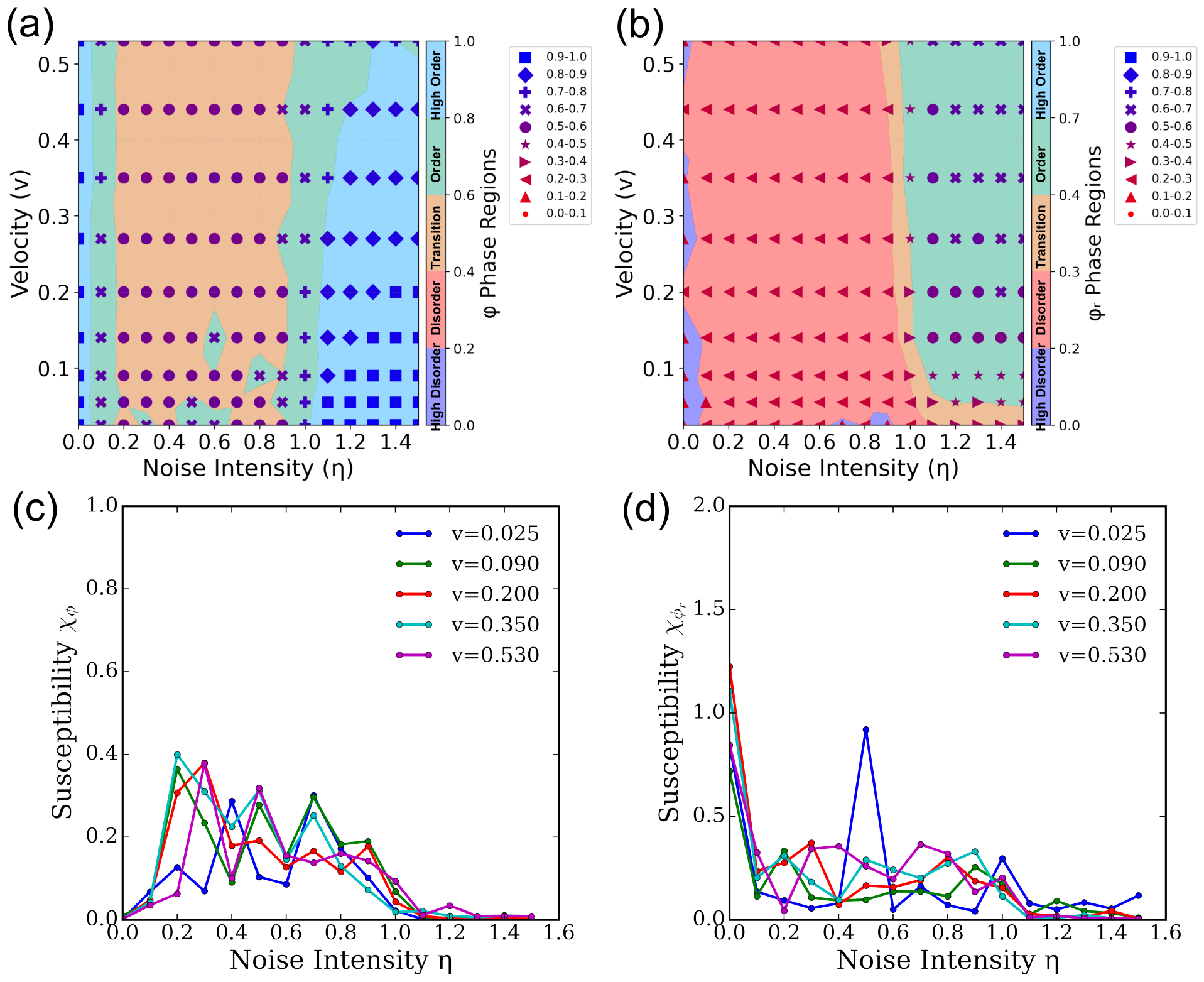}
        \caption{Phase and susceptibility analysis of Vicsek agents in a complex noise environment. (a, b) Two-dimensional heat maps of the global order parameter \(\phi\) (a) and rotational order parameter \(\phi_r\) (b) as functions of velocity (\(v\)) and noise intensity (\(\eta\)). Colors, generated using Python's Matplotlib, represent order strength, with phase regions defined as Highly Disordered (0--0.2), Disordered (0.2--0.4), Transitional (0.4--0.6), Ordered (0.6--0.8), and Highly Ordered (0.8--1.0) for \(\phi\), and 0--0.2, 0.2--0.3, 0.3--0.4, 0.4--0.7, 0.7--1.0 for \(\phi_r\), chosen for visualization clarity. (c, d) Corresponding susceptibility plots, \(\chi_\phi\) (c) and \(\chi_{\phi_r}\) (d), highlighting fluctuations in the order parameters across the phase space.}
    	\label{ffigure2}
\end{figure*}

Complex noise scenario and the characteristic dynamics of the Vicsek agents used in this work is illustrated in Figure~\ref{ffigure1}, characterized by a noiseless circular region ($\eta_{\text{c}} = 0.0$, radius $R = 12$) surrounded by a noisy outer region ($\eta_{\text{b}} \leq 1.5$). Panel (a) shows the schematic, with blue agents inside the circle aligning without noise and red agents outside experiencing tunable noise, interacting within radius $r_c$. Panel (b) captures a simulation snapshot of a flock of $N=30$ agents interacting with the noise interface capturing both the perpendicular approach and a tilted approach. Initially ($t=0$), the flock approaches perpendicularly. By $t=32$, outer agents encounter the noisy region, causing turbulence. At $t=42$, the flock tilts along the interface, and by $t=109$, noise-driven turbulence induces an inward circling motion, highlighting the emergence of rotational order.

{

Figure~\ref{ffigure2} illustrates the phase behavior and susceptibility of Vicsek agents ($N=300$) in a complex noise environment. Panels (a) and (b) present two-dimensional heat maps of the global order parameter \(\phi\) and rotational order parameter \(\phi_r\), respectively, as functions of velocity (\(v = 0.025\) to 0.530) and noise intensity (\(\eta_{\text{b}} = 0.0\) to 1.5). At low velocity (\(v = 0.025\)) and low noise (\(\eta_{\text{b}} = 0.0\)), \(\phi \approx 0.965\) indicates a highly ordered phase, while \(\phi_r \approx 0.144\) reflects weak rotational order. As \(\eta_{\text{b}}\) increases, \(\phi\) decreases to a minimum of \(\sim 0.576\) at \(\eta_{\text{b}} = 0.9\), then rises to \(\sim 0.960\) at \(\eta_{\text{b}} = 1.5\), suggesting a re-entrant ordered phase driven by noise-induced confinement. This re-entrant transition occurs at $\eta_b \sim 1.0$.
In contrast, \(\phi_r\) increases monotonically with noise, peaking at \(\sim 0.353\) at \(\eta_{\text{b}} = 1.5\) for low velocities and up to \(\sim 0.665\) at higher velocities (\(v = 0.530\)), reflecting enhanced rotational dynamics due to noise-driven turbulence at the interface. The phase regions, defined for visualization (Highly Disordered: \(\phi = 0\)--0.2, Disordered: 0.2--0.4, Transitional: 0.4--0.6, Ordered: 0.6--0.8, Highly Ordered: 0.8--1.0 for \(\phi\); 0--0.2, 0.2--0.3, 0.3--0.4, 0.4--0.7, 0.7--1.0 for \(\phi_r\)), capture these transitions. Panels (c) and (d) show the corresponding susceptibilities, \(\chi_\phi\) and \(\chi_{\phi_r}\), with low \(\chi_\phi \approx 0.000086\) at \(v = 0.025\), \(\eta_{\text{b}} = 0.0\) indicating stable alignment, and higher \(\chi_{\phi_r} \approx 1.431\) reflecting rotational fluctuations, particularly in the transitional region (\(\phi \approx 0.4\)--0.6, \(\phi_r \approx 0.3\)--0.4). For a given value of velocity, `U' shaped phase plot for $\phi$ and an increasing $\phi_r$ represents the characteristic re-entrant phase of global ordering in the sudden annealed interface, purely emerging due to flock rotation inside the noiseless region to avoid the strong turbulence outside the circle. These trends underscore the role of environmental noise in modulating order and fluctuations, with high noise promoting rotational order, global order and confinement within the noiseless region.
}

{

\begin{figure}[t]
    	\includegraphics[width=.78\textwidth]{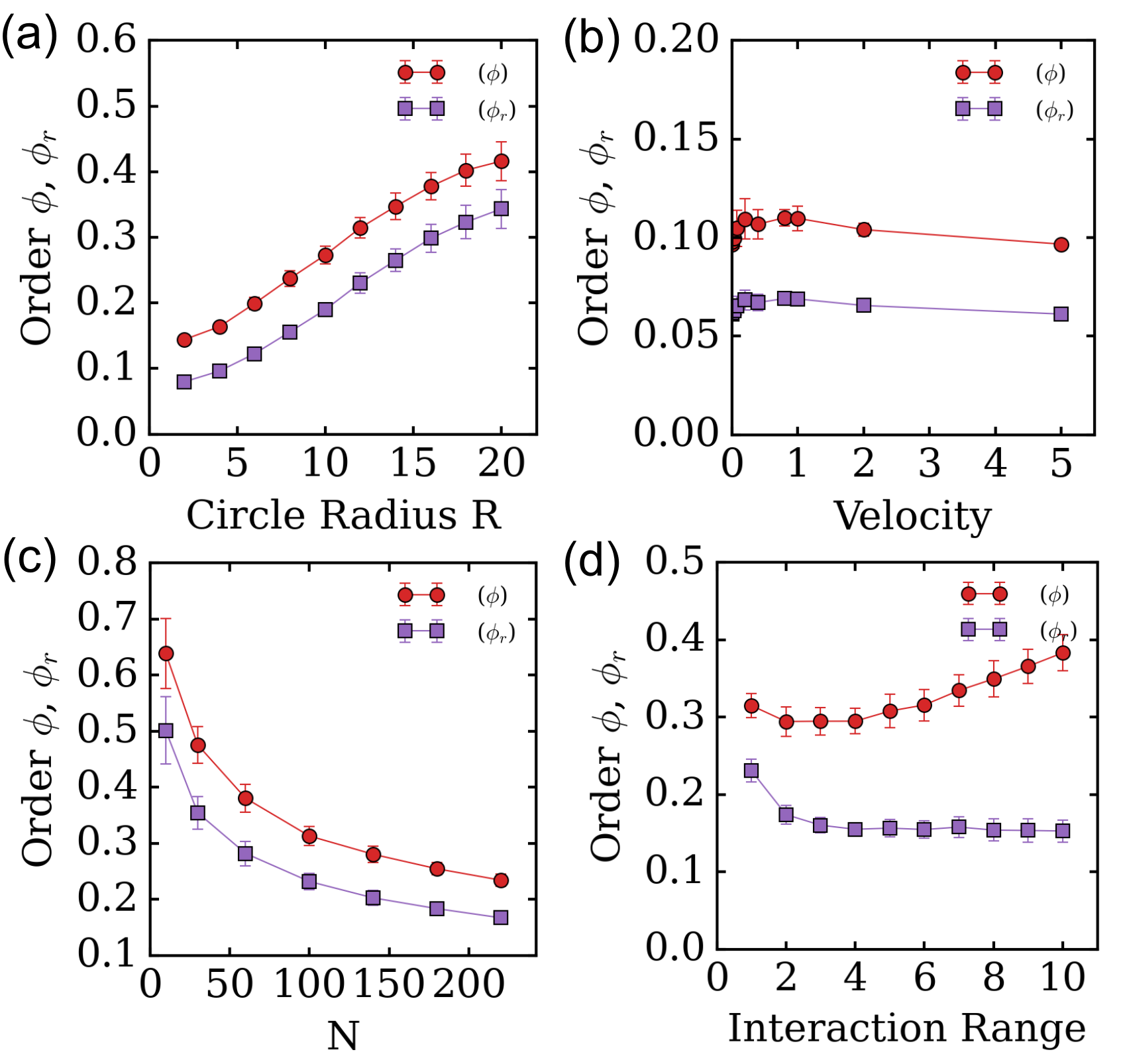}
        \caption{Order parameters \(\phi\) and \(\phi_r\) across various system parameters in a complex noise environment. (a) Variation with boundary distance \(R\), the radius of the noiseless circular region. (b) Variation with a gradual noise gradient, increasing with the distance \(r_{i,c}\) of agents from the box center. (c) Variation with particle number \(N\). (d) Variation with interaction range \(r_c\).}
    	\label{ffigure3}
        \end{figure}

Figure~\ref{ffigure3} illustrates the sensitivity of the global order parameter \(\phi\) and rotational order parameter \(\phi_r\) to key system parameters in a modified Vicsek model with $N=100$, $v=0.5$, \(R=12\) and \(\eta_{\text{b}} = 1.0\). Panel (a) shows that increasing the boundary distance \(R\) from 2.0 to 20.0 enhances \(\phi\) from 0.1440 to 0.4161 and \(\phi_r\) from 0.0797 to 0.3432, reflecting stronger global and rotational order as more agents experience the noiseless region, promoting alignment and inward circling. Panel (b) examines a gradual noise gradient, increasing with the distance \(r_{i,c}\) from the box center (noise from 0.004 to 5.0). Both \(\phi\) and \(\phi_r\) peak at intermediate noise (0.1101 and 0.0690 at noise = 0.8), suggesting an optimal noise level where turbulence enhances order before high noise induces disorder. This plot also undescores the pivotal role of sudden annealing over a gradual noise increment, and suggests gradual noise increment induces disorder, whereas sudden annealing induces re-entrant order in flocking.
Panel (c) reveals that increasing the particle number \(N\) from 10 to 220 reduces \(\phi\) from 0.6386 to 0.2340 and \(\phi_r\) from 0.5011 to 0.1668, indicating that higher densities disrupts order due to crowding and flock pressure that builds at the virtual interface. Panel (d) shows that increasing the interaction range \(r_c\) from 1.0 to 10.0 raises \(\phi\) from 0.3149 to 0.3832, enhancing global alignment, but decreases \(\phi_r\) from 0.2307 to 0.1527, suggesting that larger interaction ranges favor linear motion over rotational dynamics. These trends highlight the interplay of environmental and system parameters in modulating collective behavior, with larger noiseless regions and smaller particle numbers promoting order, while noise gradients and interaction ranges reveal competing effects on global and rotational order.

}

{

\begin{figure*}[t]
    	\includegraphics[width=.88\textwidth]{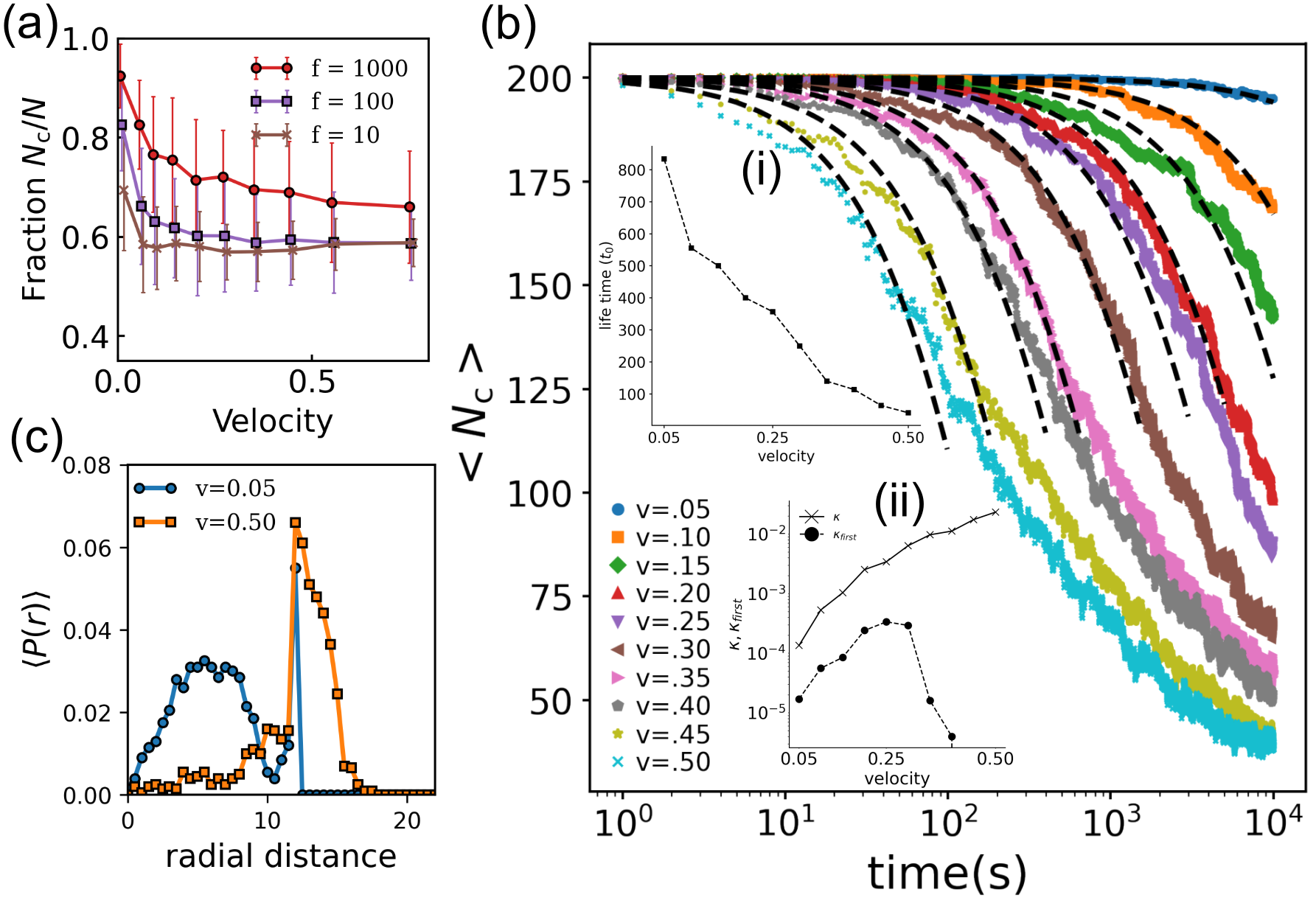}
    	\caption{Dynamics of Vicsek agents ($N=200$) in a complex noise environment. (a) Survival rate \(N_{\text{c}}/N\) averaged over 100 ensembles, showing the fraction of particles escaping the noiseless circular region per attempt frequency \(f = 1/\Delta t\), as a function of time for a perpendicular approach of first visit to the noise interface. (b) Ensemble average probability distribution \(P(r)\) of finding agents with a bimodal velocity distribution in the complex environment, recorded at \(t = 10^4\). (c) Ensemble average number of particles \(N_{\text{c}}\) in the noiseless region over time, with a dotted line representing an exponential fit \(N_{\text{c}} = N \exp(-t/t_0)\). Inset (i) shows the lifetime \(t_0\) as a function of velocity \(v\). Inset (ii) plots the steady-state escape rates \(\kappa\) and \(\kappa_{\text{first}}\) versus velocity \(v\).}
    	\label{ffigure4}
        \end{figure*}

Figure~\ref{ffigure4} illustrates the survial and retention dynamics with \(R = 16\)) surrounded, \(\eta_{\text{b}} = 1.0\), and $N=200$. Panel (a) shows the survival rate \(N_{\text{c}}/N\), the fraction of particles left in the noiseless region per attempt frequency \(f = 1/\Delta t\), defined as the number of noisy instances particles attempt per unit time when close to the interface, averaged over 100 ensembles for a perpendicular approach to the noise interface. 
At attempt frequency $f=1000$, \(N_{\text{c}}/N\) decreases from 0.9244 at \(v = 0.005\) to 0.6602 at \(v = 0.75\), indicating stronger retention at lower velocities, with higher variability (standard deviation up to 0.1272) at intermediate velocities. As the attempt frequency is lowered to $f=10$, \(N_{\text{c}}/N\) from 0.69 at \(v = 0.005\) to 0.59 at \(v = 0.75\), signifying the effect of the number of attempt agent interacts with the interface before moving far from the interface.

Panel (b) presents the probability distribution \(P(r)\) of agents with a bi-motility mixture ($N_1=50$,$v_1=0.05$ and $N_2=50$,$v_2=1.0$ ) at \(t = 10^4\). The slow moving particles, $P_1(r)$ shows a bimodal peak, both inside the noiseless region near $R\sim 5$ and $R \sim 12$. Whereas, the fast moving agents $v=1.0$, escape the complex noise trap and shows a single peak outside the interface $R\sim13$ indicating the segregation of agents based on the noisy trap induced by turbulence-driven circling. 

Panel (c) shows the average number of particles \(N_{\text{c}}\) in the noiseless region over time, exhibiting a bi-phasic decay: an initial exponential decay (\(N_{\text{c}} = N \exp(-t/t_0)\)), which slows over time as fewer particles remain. Inset (i) plots the lifetime \(t_0\), decreasing from 833.33 at \(v = 0.05\) to 41.66 at \(v = 0.50\), suggesting longer retention at slower velocities. Inset (ii) shows steady-state escape rates \(\kappa\) (increasing from 0.000135 at \(v = 0.05\) to 0.023344 at \(v = 0.50\)) and \(\kappa_{\text{first}}\) (peaking at 0.000336 at \(v = 0.25\)), indicating that higher velocities drive more escapes, with first-passage events maximized at intermediate velocities. These results highlight the interplay of velocity and noise in modulating confinement and escape, with bimodal velocity distributions enhancing rotational dynamics near the interface.
        
}

\section{Conclusion  $\&$ Discussion}\label{conc}
This study explores the collective dynamics of Vicsek agents in a complex noise environment, characterized by a noiseless circular region (\(\eta_{\text{circle}} = 0.0\), radius \(R\)) surrounded by a noisy outer region (\(\eta_{\text{box}} \leq 1.5\)), with mutually repelling interactions (\(d_{\text{min}} = 0.1\)). Our findings reveal how environmental heterogeneity (sudden annealing) modulates flocking and swarming behaviors, with significant implications for understanding and controlling active matter systems.

The phase behavior, illustrated in Figure~\ref{ffigure2}, shows that the global order parameter \(\phi\) exhibits a U-shaped trend with increasing noise intensity (\(\eta_{\text{b}}\)), dropping to a minimum of \(\sim 0.576\) at \(\eta_{\text{b}} = 0.9\) and recovering to \(\sim 0.960\) at \(\eta_{\text{b}} = 1.5\) for low velocity (\(v = 0.005\)). In contrast, the rotational order parameter \(\phi_r\) increases monotonically, peaking at \(\sim 0.665\) for higher velocities (\(v = 0.440\), \(\eta_{\text{b}} = 1.4\)). These trends, supported by susceptibility plots (\(\chi_\phi \approx 0.000086\), \(\chi_{\phi_r} \approx 1.431\) at \(v = 0.005\), \(\eta_{\text{b}} = 0.0\)), indicate that low noise promotes global alignment, while high noise enhances rotational dynamics which in turn results in re-entrant global ordering, due to turbulence at the noise interface driving inward circling, as observed in simulation snapshots, see Video-1.

Figure~\ref{ffigure3} further demonstrates the sensitivity of \(\phi\) and \(\phi_r\) to system parameters. Increasing the noiseless region’s radius \(R\) from 2.0 to 20.0 boosts \(\phi\) (0.1440 to 0.4161) and \(\phi_r\) (0.0797 to 0.3432), reflecting enhanced order as more agents experience noise-free alignment. A gradual noise gradient, increasing with distance \(r_{i,c}\) from the box center, yields a peak in order (\(\phi \approx 0.1101\), \(\phi_r \approx 0.0690\) at noise = 0.8), suggesting an optimal noise level for balancing confinement and turbulence, and portraying the impact of sudden annealing over gradual annealing in noise levels. Conversely, increasing particle number \(N\) from 10 to 220 reduces \(\phi\) (0.6386 to 0.2340) and \(\phi_r\) (0.5011 to 0.1668), due to crowding and flock pressure at the interface. Larger interaction ranges \(r_c\) (1.0 to 10.0) enhance \(\phi\) (0.3149 to 0.3832) but reduce \(\phi_r\) (0.2307 to 0.1527), indicating a trade-off between global alignment and rotational motion.

Survival kinetics, shown in Figure~\ref{ffigure4}, reveal that the fraction of particles survived in the noiseless region (\(N_{\text{c}}/N\)) decreases with velocity (0.9244 at \(v = 0.005\) to 0.6602 at \(v = 0.75\) for \(\Delta t = 0.001\)), with higher velocities increasing steady-state escape rates (\(\kappa\) from 0.000135 to 0.023344). First-passage escape rates (\(\kappa_{\text{first}}\)) peak at intermediate velocities (0.000336 at \(v = 0.25\)), reflecting a balance between mobility and confinement. The probability distribution \(P(r)\) at \(t = 10^4\) shows bimotility mixture of agents shows a segregation with slower moving particles $v=0.05$, $P_1(r)$ peaks inside the noiseless region, whereas fast moving particles $P_2(r)$ peaks outside the noiseless region. The number of particles in the noiseless region (\(N_{\text{c}}\)) follows a bi-phasic decay, initially exponential (\(N_{\text{c}} = N \exp(-t/t_0)\)) then slowing, with lifetime \(t_0\) decreasing with velocity (833.33 to 41.66).

These results underscore the role of environmental heterogeneity in driving emergent behaviors. The re-entrant global order at high noise suggests that strong turbulence confines agents within the noiseless region, enhancing \(\phi\), while rotational order peaks due to interface-driven circling, consistent with natural systems like fish schools navigating turbulent waters or bacterial swarms in heterogeneous media. The sensitivity to \(R\), \(N\), and \(r_c\) highlights how system parameters can be tuned to control collective dynamics, with smaller \(N\) and larger \(R\) favoring order, and intermediate noise gradients optimizing both \(\phi\) and \(\phi_r\). The escape dynamics and bi-motility agents segregation suggest trapping mechanism for synthetic swarms, where velocity gradients and noise interfaces could guide robotic collectives.

Future work could explore the functional form of the noise gradient (\(\eta(r_{i,c})\)), investigate the lifetime scaling anomaly, and extend the model to three dimensions or non-uniform velocity profiles. Specially for active colloids, magnetic field can be tuned to create strong noisy environment and increasing the cohesion of chemically propelled colloids. These findings provide a robust framework for predicting and manipulating active agent behavior in complex environments, bridging biological and synthetic systems.

\section*{Competing Interests}
The author declares no competing interests. This work was conducted as part of the PhD work of Mohd Yasir Khan and therefore should be considered as part of the PhD thesis.

\section*{Data Availability}
The data that support the findings of this study are available within the article.

\section*{Acknowledgments}
Mohd Yasir Khan would like to thank Prof. Wang Wei (HIT-Shenzhen), Prof. Sujin Babu (IIT-Delhi) and Dr. Mihir Durve for input and critical insights of flocking kinetics. Mohd Yasir Khan would like to thank University Grants Commission(UGC) for the PhD fellowship.

\cleardoublepage
	
	\newpage

	\vspace*{1\baselineskip}
	\bibliographystyle{apsrev4-2} 
		\nocite{*}
	\bibliography{0main.bib} 

\begin{thebibliography}{27}%
\makeatletter
\providecommand \@ifxundefined [1]{%
 \@ifx{#1\undefined}
}%
\providecommand \@ifnum [1]{%
 \ifnum #1\expandafter \@firstoftwo
 \else \expandafter \@secondoftwo
 \fi
}%
\providecommand \@ifx [1]{%
 \ifx #1\expandafter \@firstoftwo
 \else \expandafter \@secondoftwo
 \fi
}%
\providecommand \natexlab [1]{#1}%
\providecommand \enquote  [1]{``#1''}%
\providecommand \bibnamefont  [1]{#1}%
\providecommand \bibfnamefont [1]{#1}%
\providecommand \citenamefont [1]{#1}%
\providecommand \href@noop [0]{\@secondoftwo}%
\providecommand \href [0]{\begingroup \@sanitize@url \@href}%
\providecommand \@href[1]{\@@startlink{#1}\@@href}%
\providecommand \@@href[1]{\endgroup#1\@@endlink}%
\providecommand \@sanitize@url [0]{\catcode `\\12\catcode `\$12\catcode `\&12\catcode `\#12\catcode `\^12\catcode `\_12\catcode `\%12\relax}%
\providecommand \@@startlink[1]{}%
\providecommand \@@endlink[0]{}%
\providecommand \url  [0]{\begingroup\@sanitize@url \@url }%
\providecommand \@url [1]{\endgroup\@href {#1}{\urlprefix }}%
\providecommand \urlprefix  [0]{URL }%
\providecommand \Eprint [0]{\href }%
\providecommand \doibase [0]{https://doi.org/}%
\providecommand \selectlanguage [0]{\@gobble}%
\providecommand \bibinfo  [0]{\@secondoftwo}%
\providecommand \bibfield  [0]{\@secondoftwo}%
\providecommand \translation [1]{[#1]}%
\providecommand \BibitemOpen [0]{}%
\providecommand \bibitemStop [0]{}%
\providecommand \bibitemNoStop [0]{.\EOS\space}%
\providecommand \EOS [0]{\spacefactor3000\relax}%
\providecommand \BibitemShut  [1]{\csname bibitem#1\endcsname}%
\let\auto@bib@innerbib\@empty
\bibitem [{\citenamefont {Chatterjee}\ \emph {et~al.}(2025)\citenamefont {Chatterjee}, \citenamefont {Chakrabortty},\ and\ \citenamefont {Bhamla}}]{chatterjee2025noisy}%
  \BibitemOpen
  \bibfield  {author} {\bibinfo {author} {\bibfnamefont {A.}~\bibnamefont {Chatterjee}}, \bibinfo {author} {\bibfnamefont {T.}~\bibnamefont {Chakrabortty}},\ and\ \bibinfo {author} {\bibfnamefont {S.}~\bibnamefont {Bhamla}},\ }\href@noop {} {\bibfield  {journal} {\bibinfo  {journal} {arXiv preprint arXiv:2508.16031}\ } (\bibinfo {year} {2025})}\BibitemShut {NoStop}%
\bibitem [{\citenamefont {Frey}\ and\ \citenamefont {Kroy}(2005)}]{frey2005brownian}%
  \BibitemOpen
  \bibfield  {author} {\bibinfo {author} {\bibfnamefont {E.}~\bibnamefont {Frey}}\ and\ \bibinfo {author} {\bibfnamefont {K.}~\bibnamefont {Kroy}},\ }\href@noop {} {\bibfield  {journal} {\bibinfo  {journal} {Annalen der Physik}\ }\textbf {\bibinfo {volume} {517}},\ \bibinfo {pages} {20} (\bibinfo {year} {2005})}\BibitemShut {NoStop}%
\bibitem [{\citenamefont {Ramaswamy}(2010)}]{ramaswamy2010mechanics}%
  \BibitemOpen
  \bibfield  {author} {\bibinfo {author} {\bibfnamefont {S.}~\bibnamefont {Ramaswamy}},\ }\href@noop {} {\bibfield  {journal} {\bibinfo  {journal} {Annu. Rev. Condens. Matter Phys.}\ }\textbf {\bibinfo {volume} {1}},\ \bibinfo {pages} {323} (\bibinfo {year} {2010})}\BibitemShut {NoStop}%
\bibitem [{\citenamefont {Jhawar}\ \emph {et~al.}(2020)\citenamefont {Jhawar}, \citenamefont {Morris}, \citenamefont {Amith-Kumar}, \citenamefont {Danny~Raj}, \citenamefont {Rogers}, \citenamefont {Rajendran},\ and\ \citenamefont {Guttal}}]{jhawar2020noise}%
  \BibitemOpen
  \bibfield  {author} {\bibinfo {author} {\bibfnamefont {J.}~\bibnamefont {Jhawar}}, \bibinfo {author} {\bibfnamefont {R.~G.}\ \bibnamefont {Morris}}, \bibinfo {author} {\bibfnamefont {U.}~\bibnamefont {Amith-Kumar}}, \bibinfo {author} {\bibfnamefont {M.}~\bibnamefont {Danny~Raj}}, \bibinfo {author} {\bibfnamefont {T.}~\bibnamefont {Rogers}}, \bibinfo {author} {\bibfnamefont {H.}~\bibnamefont {Rajendran}},\ and\ \bibinfo {author} {\bibfnamefont {V.}~\bibnamefont {Guttal}},\ }\href@noop {} {\bibfield  {journal} {\bibinfo  {journal} {Nature Physics}\ }\textbf {\bibinfo {volume} {16}},\ \bibinfo {pages} {488} (\bibinfo {year} {2020})}\BibitemShut {NoStop}%
\bibitem [{\citenamefont {Yatesa}\ \emph {et~al.}(2009)\citenamefont {Yatesa}, \citenamefont {Erbana}, \citenamefont {Escuderoc}, \citenamefont {Couzind}, \citenamefont {Buhle}, \citenamefont {Kevrekidisf}, \citenamefont {Mainia},\ and\ \citenamefont {Sumpterh}}]{yatesa2009inherent}%
  \BibitemOpen
  \bibfield  {author} {\bibinfo {author} {\bibfnamefont {C.~A.}\ \bibnamefont {Yatesa}}, \bibinfo {author} {\bibfnamefont {R.}~\bibnamefont {Erbana}}, \bibinfo {author} {\bibfnamefont {C.}~\bibnamefont {Escuderoc}}, \bibinfo {author} {\bibfnamefont {I.~D.}\ \bibnamefont {Couzind}}, \bibinfo {author} {\bibfnamefont {J.}~\bibnamefont {Buhle}}, \bibinfo {author} {\bibfnamefont {I.~G.}\ \bibnamefont {Kevrekidisf}}, \bibinfo {author} {\bibfnamefont {P.~K.}\ \bibnamefont {Mainia}},\ and\ \bibinfo {author} {\bibfnamefont {D.~J.}\ \bibnamefont {Sumpterh}},\ }\href@noop {} {\bibfield  {journal} {\bibinfo  {journal} {PNAS}\ }\textbf {\bibinfo {volume} {106}} (\bibinfo {year} {2009})}\BibitemShut {NoStop}%
\bibitem [{\citenamefont {Yates}\ \emph {et~al.}(2009)\citenamefont {Yates}, \citenamefont {Erban}, \citenamefont {Escudero}, \citenamefont {Couzin}, \citenamefont {Buhl}, \citenamefont {Kevrekidis}, \citenamefont {Maini},\ and\ \citenamefont {Sumpter}}]{yates2009united}%
  \BibitemOpen
  \bibfield  {author} {\bibinfo {author} {\bibfnamefont {C.~A.}\ \bibnamefont {Yates}}, \bibinfo {author} {\bibfnamefont {R.}~\bibnamefont {Erban}}, \bibinfo {author} {\bibfnamefont {C.}~\bibnamefont {Escudero}}, \bibinfo {author} {\bibfnamefont {I.~D.}\ \bibnamefont {Couzin}}, \bibinfo {author} {\bibfnamefont {J.}~\bibnamefont {Buhl}}, \bibinfo {author} {\bibfnamefont {I.~G.}\ \bibnamefont {Kevrekidis}}, \bibinfo {author} {\bibfnamefont {P.~K.}\ \bibnamefont {Maini}},\ and\ \bibinfo {author} {\bibfnamefont {D.}~\bibnamefont {Sumpter}},\ }\href@noop {} {\bibfield  {journal} {\bibinfo  {journal} {Proceedings of the National Academy of Sciences}\ }\textbf {\bibinfo {volume} {106}},\ \bibinfo {pages} {5464} (\bibinfo {year} {2009})}\BibitemShut {NoStop}%
\bibitem [{\citenamefont {Couzin}\ \emph {et~al.}(2005)\citenamefont {Couzin}, \citenamefont {Krause}, \citenamefont {Franks},\ and\ \citenamefont {Levin}}]{couzin2005effective}%
  \BibitemOpen
  \bibfield  {author} {\bibinfo {author} {\bibfnamefont {I.~D.}\ \bibnamefont {Couzin}}, \bibinfo {author} {\bibfnamefont {J.}~\bibnamefont {Krause}}, \bibinfo {author} {\bibfnamefont {N.~R.}\ \bibnamefont {Franks}},\ and\ \bibinfo {author} {\bibfnamefont {S.~A.}\ \bibnamefont {Levin}},\ }\href@noop {} {\bibfield  {journal} {\bibinfo  {journal} {Nature}\ }\textbf {\bibinfo {volume} {433}},\ \bibinfo {pages} {513} (\bibinfo {year} {2005})}\BibitemShut {NoStop}%
\bibitem [{\citenamefont {King}\ and\ \citenamefont {Sumpter}(2012)}]{king2012murmurations}%
  \BibitemOpen
  \bibfield  {author} {\bibinfo {author} {\bibfnamefont {A.~J.}\ \bibnamefont {King}}\ and\ \bibinfo {author} {\bibfnamefont {D.~J.}\ \bibnamefont {Sumpter}},\ }\href@noop {} {\bibfield  {journal} {\bibinfo  {journal} {Current Biology}\ }\textbf {\bibinfo {volume} {22}},\ \bibinfo {pages} {R112} (\bibinfo {year} {2012})}\BibitemShut {NoStop}%
\bibitem [{\citenamefont {Mathijssen}\ \emph {et~al.}(2019)\citenamefont {Mathijssen}, \citenamefont {Culver}, \citenamefont {Bhamla},\ and\ \citenamefont {Prakash}}]{mathijssen2019collective}%
  \BibitemOpen
  \bibfield  {author} {\bibinfo {author} {\bibfnamefont {A.~J.}\ \bibnamefont {Mathijssen}}, \bibinfo {author} {\bibfnamefont {J.}~\bibnamefont {Culver}}, \bibinfo {author} {\bibfnamefont {M.~S.}\ \bibnamefont {Bhamla}},\ and\ \bibinfo {author} {\bibfnamefont {M.}~\bibnamefont {Prakash}},\ }\href@noop {} {\bibfield  {journal} {\bibinfo  {journal} {Nature}\ }\textbf {\bibinfo {volume} {571}},\ \bibinfo {pages} {560} (\bibinfo {year} {2019})}\BibitemShut {NoStop}%
\bibitem [{\citenamefont {Peleg}\ \emph {et~al.}(2018)\citenamefont {Peleg}, \citenamefont {Peters}, \citenamefont {Salcedo},\ and\ \citenamefont {Mahadevan}}]{peleg2018collective}%
  \BibitemOpen
  \bibfield  {author} {\bibinfo {author} {\bibfnamefont {O.}~\bibnamefont {Peleg}}, \bibinfo {author} {\bibfnamefont {J.~M.}\ \bibnamefont {Peters}}, \bibinfo {author} {\bibfnamefont {M.~K.}\ \bibnamefont {Salcedo}},\ and\ \bibinfo {author} {\bibfnamefont {L.}~\bibnamefont {Mahadevan}},\ }\href@noop {} {\bibfield  {journal} {\bibinfo  {journal} {Nature Physics}\ }\textbf {\bibinfo {volume} {14}},\ \bibinfo {pages} {1193} (\bibinfo {year} {2018})}\BibitemShut {NoStop}%
\bibitem [{\citenamefont {Zhang}\ and\ \citenamefont {Shao}(2021)}]{zhang2021collective}%
  \BibitemOpen
  \bibfield  {author} {\bibinfo {author} {\bibfnamefont {B.-Q.}\ \bibnamefont {Zhang}}\ and\ \bibinfo {author} {\bibfnamefont {Z.-G.}\ \bibnamefont {Shao}},\ }\href@noop {} {\bibfield  {journal} {\bibinfo  {journal} {Physica A: Statistical Mechanics and its Applications}\ }\textbf {\bibinfo {volume} {563}},\ \bibinfo {pages} {125382} (\bibinfo {year} {2021})}\BibitemShut {NoStop}%
\bibitem [{\citenamefont {Khan}\ and\ \citenamefont {Babu}(2022)}]{khan2022effect}%
  \BibitemOpen
  \bibfield  {author} {\bibinfo {author} {\bibfnamefont {M.~Y.}\ \bibnamefont {Khan}}\ and\ \bibinfo {author} {\bibfnamefont {S.~B.}\ \bibnamefont {Babu}},\ }\href {https://doi.org/10.1063/5.0090619} {\bibfield  {journal} {\bibinfo  {journal} {Physics of Fluids}\ }\textbf {\bibinfo {volume} {34}},\ \bibinfo {pages} {061901} (\bibinfo {year} {2022})},\ \Eprint {https://arxiv.org/abs/https://pubs.aip.org/aip/pof/article-pdf/doi/10.1063/5.0090619/16565809/061901\_1\_online.pdf} {https://pubs.aip.org/aip/pof/article-pdf/doi/10.1063/5.0090619/16565809/061901\_1\_online.pdf} \BibitemShut {NoStop}%
\bibitem [{\citenamefont {Peng}\ \emph {et~al.}(2025)\citenamefont {Peng}, \citenamefont {Yasir~Khan}, \citenamefont {Gao},\ and\ \citenamefont {Wang}}]{peng2025self}%
  \BibitemOpen
  \bibfield  {author} {\bibinfo {author} {\bibfnamefont {Y.}~\bibnamefont {Peng}}, \bibinfo {author} {\bibfnamefont {M.}~\bibnamefont {Yasir~Khan}}, \bibinfo {author} {\bibfnamefont {Y.}~\bibnamefont {Gao}},\ and\ \bibinfo {author} {\bibfnamefont {W.}~\bibnamefont {Wang}},\ }\href {https://doi.org/https://doi.org/10.1002/asia.202400923} {\bibfield  {journal} {\bibinfo  {journal} {Chemistry – An Asian Journal}\ }\textbf {\bibinfo {volume} {20}},\ \bibinfo {pages} {e202400923} (\bibinfo {year} {2025})},\ \Eprint {https://arxiv.org/abs/https://aces.onlinelibrary.wiley.com/doi/pdf/10.1002/asia.202400923} {https://aces.onlinelibrary.wiley.com/doi/pdf/10.1002/asia.202400923} \BibitemShut {NoStop}%
\bibitem [{\citenamefont {Mackay}\ \emph {et~al.}(2025)\citenamefont {Mackay}, \citenamefont {Janzen}, \citenamefont {Fernandez},\ and\ \citenamefont {Sknepnek}}]{mackay2025emergent}%
  \BibitemOpen
  \bibfield  {author} {\bibinfo {author} {\bibfnamefont {E.~D.}\ \bibnamefont {Mackay}}, \bibinfo {author} {\bibfnamefont {G.}~\bibnamefont {Janzen}}, \bibinfo {author} {\bibfnamefont {D.}~\bibnamefont {Fernandez}},\ and\ \bibinfo {author} {\bibfnamefont {R.}~\bibnamefont {Sknepnek}},\ }\href@noop {} {\bibfield  {journal} {\bibinfo  {journal} {arXiv preprint arXiv:2505.24730}\ } (\bibinfo {year} {2025})}\BibitemShut {NoStop}%
\bibitem [{\citenamefont {Xiao}\ \emph {et~al.}(2018)\citenamefont {Xiao}, \citenamefont {Wei},\ and\ \citenamefont {Wang}}]{xiao2018review}%
  \BibitemOpen
  \bibfield  {author} {\bibinfo {author} {\bibfnamefont {Z.}~\bibnamefont {Xiao}}, \bibinfo {author} {\bibfnamefont {M.}~\bibnamefont {Wei}},\ and\ \bibinfo {author} {\bibfnamefont {W.}~\bibnamefont {Wang}},\ }\href@noop {} {\bibfield  {journal} {\bibinfo  {journal} {ACS applied materials \& interfaces}\ }\textbf {\bibinfo {volume} {11}},\ \bibinfo {pages} {6667} (\bibinfo {year} {2018})}\BibitemShut {NoStop}%
\bibitem [{\citenamefont {Varma}\ \emph {et~al.}(2024)\citenamefont {Varma}, \citenamefont {Jaglan}, \citenamefont {Khan},\ and\ \citenamefont {Babu}}]{varma2024breaking}%
  \BibitemOpen
  \bibfield  {author} {\bibinfo {author} {\bibfnamefont {V.~A.}\ \bibnamefont {Varma}}, \bibinfo {author} {\bibfnamefont {S.}~\bibnamefont {Jaglan}}, \bibinfo {author} {\bibfnamefont {M.~Y.}\ \bibnamefont {Khan}},\ and\ \bibinfo {author} {\bibfnamefont {S.~B.}\ \bibnamefont {Babu}},\ }\href {https://doi.org/10.1039/D3CP03681B} {\bibfield  {journal} {\bibinfo  {journal} {Phys. Chem. Chem. Phys.}\ }\textbf {\bibinfo {volume} {26}},\ \bibinfo {pages} {1385} (\bibinfo {year} {2024})}\BibitemShut {NoStop}%
\bibitem [{\citenamefont {L{\"o}wen}(2025)}]{lowen2025colloidal}%
  \BibitemOpen
  \bibfield  {author} {\bibinfo {author} {\bibfnamefont {H.}~\bibnamefont {L{\"o}wen}},\ }\href@noop {} {\bibfield  {journal} {\bibinfo  {journal} {Newton}\ } (\bibinfo {year} {2025})}\BibitemShut {NoStop}%
\bibitem [{\citenamefont {Evers}\ \emph {et~al.}(2013)\citenamefont {Evers}, \citenamefont {Hanes}, \citenamefont {Zunke}, \citenamefont {Capellmann}, \citenamefont {Bewerunge}, \citenamefont {Dalle-Ferrier}, \citenamefont {Jenkins}, \citenamefont {Ladadwa}, \citenamefont {Heuer}, \citenamefont {Castaneda-Priego} \emph {et~al.}}]{evers2013colloids}%
  \BibitemOpen
  \bibfield  {author} {\bibinfo {author} {\bibfnamefont {F.}~\bibnamefont {Evers}}, \bibinfo {author} {\bibfnamefont {R.}~\bibnamefont {Hanes}}, \bibinfo {author} {\bibfnamefont {C.}~\bibnamefont {Zunke}}, \bibinfo {author} {\bibfnamefont {R.}~\bibnamefont {Capellmann}}, \bibinfo {author} {\bibfnamefont {J.}~\bibnamefont {Bewerunge}}, \bibinfo {author} {\bibfnamefont {C.}~\bibnamefont {Dalle-Ferrier}}, \bibinfo {author} {\bibfnamefont {M.}~\bibnamefont {Jenkins}}, \bibinfo {author} {\bibfnamefont {I.}~\bibnamefont {Ladadwa}}, \bibinfo {author} {\bibfnamefont {A.}~\bibnamefont {Heuer}}, \bibinfo {author} {\bibfnamefont {R.}~\bibnamefont {Castaneda-Priego}}, \emph {et~al.},\ }\href@noop {} {\bibfield  {journal} {\bibinfo  {journal} {The European Physical Journal Special Topics}\ }\textbf {\bibinfo {volume} {222}},\ \bibinfo {pages} {2995} (\bibinfo {year} {2013})}\BibitemShut {NoStop}%
\bibitem [{\citenamefont {Frechette}\ \emph {et~al.}(2025)\citenamefont {Frechette}, \citenamefont {Baskaran},\ and\ \citenamefont {Hagan}}]{frechette2025active}%
  \BibitemOpen
  \bibfield  {author} {\bibinfo {author} {\bibfnamefont {L.~B.}\ \bibnamefont {Frechette}}, \bibinfo {author} {\bibfnamefont {A.}~\bibnamefont {Baskaran}},\ and\ \bibinfo {author} {\bibfnamefont {M.~F.}\ \bibnamefont {Hagan}},\ }\href@noop {} {\bibfield  {journal} {\bibinfo  {journal} {Newton}\ } (\bibinfo {year} {2025})}\BibitemShut {NoStop}%
\bibitem [{\citenamefont {Simmchen}\ \emph {et~al.}(2024)\citenamefont {Simmchen}, \citenamefont {Uspal},\ and\ \citenamefont {Wang}}]{simmchen2024active}%
  \BibitemOpen
  \bibfield  {author} {\bibinfo {author} {\bibfnamefont {J.}~\bibnamefont {Simmchen}}, \bibinfo {author} {\bibfnamefont {W.}~\bibnamefont {Uspal}},\ and\ \bibinfo {author} {\bibfnamefont {W.}~\bibnamefont {Wang}},\ }\href@noop {} {\emph {\bibinfo {title} {Active Colloids: From Fundamentals to Frontiers}}}\ (\bibinfo  {publisher} {Royal Society of Chemistry},\ \bibinfo {year} {2024})\BibitemShut {NoStop}%
\bibitem [{\citenamefont {Couzin}\ and\ \citenamefont {Li}(2023)}]{couzin2023benefits}%
  \BibitemOpen
  \bibfield  {author} {\bibinfo {author} {\bibfnamefont {I.~D.}\ \bibnamefont {Couzin}}\ and\ \bibinfo {author} {\bibfnamefont {L.}~\bibnamefont {Li}},\ }\href@noop {} {\bibfield  {journal} {\bibinfo  {journal} {elife}\ }\textbf {\bibinfo {volume} {12}},\ \bibinfo {pages} {e86807} (\bibinfo {year} {2023})}\BibitemShut {NoStop}%
\bibitem [{\citenamefont {Chakrabortty}\ and\ \citenamefont {Bhamla}(2025)}]{chakrabortty2025controlling}%
  \BibitemOpen
  \bibfield  {author} {\bibinfo {author} {\bibfnamefont {T.}~\bibnamefont {Chakrabortty}}\ and\ \bibinfo {author} {\bibfnamefont {S.}~\bibnamefont {Bhamla}},\ }\href@noop {} {\bibfield  {journal} {\bibinfo  {journal} {ArXiv}\ ,\ \bibinfo {pages} {arXiv}} (\bibinfo {year} {2025})}\BibitemShut {NoStop}%
\bibitem [{\citenamefont {Cui}\ \emph {et~al.}(2025)\citenamefont {Cui}, \citenamefont {Khan}, \citenamefont {Chen}, \citenamefont {Yan}, \citenamefont {Liu},\ and\ \citenamefont {Wang}}]{cui2025emergent}%
  \BibitemOpen
  \bibfield  {author} {\bibinfo {author} {\bibfnamefont {D.}~\bibnamefont {Cui}}, \bibinfo {author} {\bibfnamefont {M.~Y.}\ \bibnamefont {Khan}}, \bibinfo {author} {\bibfnamefont {X.}~\bibnamefont {Chen}}, \bibinfo {author} {\bibfnamefont {Z.}~\bibnamefont {Yan}}, \bibinfo {author} {\bibfnamefont {X.}~\bibnamefont {Liu}},\ and\ \bibinfo {author} {\bibfnamefont {W.}~\bibnamefont {Wang}},\ }\href {https://doi.org/10.1039/D5SM00349K} {\bibfield  {journal} {\bibinfo  {journal} {Soft Matter}\ }\textbf {\bibinfo {volume} {21}},\ \bibinfo {pages} {6391} (\bibinfo {year} {2025})}\BibitemShut {NoStop}%
\bibitem [{\citenamefont {Vicsek}\ and\ \citenamefont {Zafeiris}(2012)}]{VICSEK201271}%
  \BibitemOpen
  \bibfield  {author} {\bibinfo {author} {\bibfnamefont {T.}~\bibnamefont {Vicsek}}\ and\ \bibinfo {author} {\bibfnamefont {A.}~\bibnamefont {Zafeiris}},\ }\href {https://doi.org/https://doi.org/10.1016/j.physrep.2012.03.004} {\bibfield  {journal} {\bibinfo  {journal} {Physics Reports}\ }\textbf {\bibinfo {volume} {517}},\ \bibinfo {pages} {71} (\bibinfo {year} {2012})},\ \bibinfo {note} {collective motion}\BibitemShut {NoStop}%
\bibitem [{\citenamefont {Chan}\ and\ \citenamefont {Kanso}(2025)}]{chan2025noise}%
  \BibitemOpen
  \bibfield  {author} {\bibinfo {author} {\bibfnamefont {A.}~\bibnamefont {Chan}}\ and\ \bibinfo {author} {\bibfnamefont {E.}~\bibnamefont {Kanso}},\ }\href@noop {} {\bibfield  {journal} {\bibinfo  {journal} {arXiv preprint arXiv:2507.16102}\ } (\bibinfo {year} {2025})}\BibitemShut {NoStop}%
\bibitem [{\citenamefont {Vicsek}\ \emph {et~al.}(1995)\citenamefont {Vicsek}, \citenamefont {Czir{\'o}k}, \citenamefont {Ben-Jacob}, \citenamefont {Cohen},\ and\ \citenamefont {Shochet}}]{vicsek1995novel}%
  \BibitemOpen
  \bibfield  {author} {\bibinfo {author} {\bibfnamefont {T.}~\bibnamefont {Vicsek}}, \bibinfo {author} {\bibfnamefont {A.}~\bibnamefont {Czir{\'o}k}}, \bibinfo {author} {\bibfnamefont {E.}~\bibnamefont {Ben-Jacob}}, \bibinfo {author} {\bibfnamefont {I.}~\bibnamefont {Cohen}},\ and\ \bibinfo {author} {\bibfnamefont {O.}~\bibnamefont {Shochet}},\ }\href@noop {} {\bibfield  {journal} {\bibinfo  {journal} {Physical review letters}\ }\textbf {\bibinfo {volume} {75}},\ \bibinfo {pages} {1226} (\bibinfo {year} {1995})}\BibitemShut {NoStop}%
\bibitem [{\citenamefont {Khan}(2023)}]{IITDThesisyasir}%
  \BibitemOpen
  \bibfield  {author} {\bibinfo {author} {\bibfnamefont {M.~Y.}\ \bibnamefont {Khan}},\ }\emph {\bibinfo {title} {Collective behaviour in non equilibrium active matter}},\ \href {http://https://libcat.iitd.ac.in/bib/244418} {Ph.D. thesis},\ \bibinfo  {school} {Department of Physics, Indian Institute of Technology Delhi}, \bibinfo {address} {New Delhi, India} (\bibinfo {year} {2023}),\ \bibinfo {note} {main finding of this article follows the work of M.Y. Khan in chapter-6 of his thesis.}\BibitemShut {Stop}%
\end{thebibliography}%
\end{document}